\documentclass[11pt,a4paper]{article}
\usepackage{longtable}
\usepackage{amsmath}
\usepackage{amssymb}
\usepackage{graphicx}
\usepackage{array}
\usepackage{lipsum}
\usepackage{bm}

\newcommand*{\bea}{\begin{eqnarray}}
\newcommand*{\eea}{\end{eqnarray}}
\newcommand*{\be}{\begin{equation}}
\newcommand*{\ee}{\end{equation}}

\newcommand{\bma}{\begin{pmatrix}}
\newcommand{\ema}{\end{pmatrix}}
\usepackage[square,comma,sort&compress,numbers]{natbib}
\title{Dynamical generation of scalar mass in two Higgs doublet model under Yukawa interactions}
\author{Tajdar Mufti \footnote{tajdar.mufti@gmail.com, tajdar.mufti@lums.edu.pk} \\ Lahore University of Management Sciences\\ Opposite Sector U, D.H.A, Lahore Cantt., 54792, Pakistan}
\begin{document}
\maketitle
\begin{abstract}
Light scalars are among the expected particles in nature. If they indeed exist, dynamical generation of masses becomes an important phenomenon to investigate in scalar interactions. A two Higgs doublet model containing two complex doublet scalar fields, conveniently called the standard model Higgs and the second Higgs fields, is studied to explore the extent of dynamical mass generation and the field propagators in the model at different cutoff values. Both Higgs fields are coupled with each other by a real singlet scalar field through a modelled Yukawa interaction. The model is studied for various renormalized masses of the second Higgs field at various coupling strength. The renormalized mass of the standard model Higgs boson is kept at $125.09$ GeV. The model has strong indication of existence of critical coupling below $10^{-3}$ GeV. The observed dynamical masses are generally within $200$ MeV. The two Higgs propagators are found to be considerably stable compared to scalar singlet propagators despite cutoff effects. No phase structure in the parameter space was observed. The model is found non-trivial.
\end{abstract}
\section{Introduction}
Dynamical mass generation (DMG) \cite{Brauner:2005hw,Maris:1999nt,Fischer:2003rp,Aguilar:2005sb,Bowman:2005vx,Aguilar:2010cn,Cloet:2013jya,Mitter:2014wpa,Binosi:2016wcx,Ayala:2006sv,Libanov:2005vu,Benes:2008ir} is a non-perturbative phenomenon by definition. Historically, it has mostly been relevant to QCD physics \cite{Cornwall:1981zr}, particularly to the lightest family of quarks due to limited capacity of QCD interactions to generate large dynamical masses. In the new physics, the phenomenon is equally relevant to masses considerably lighter than the electroweak scale. As the existence of scalar sector in nature has become a possibility after the experimental finding of the Higgs boson \cite{pdg,Carena:2002es,Aad:2012tfa,Chatrchyan:2012xdj,MalbertionbehalfoftheCMSCollaboration:2018eqs}, ascertaining the extent of dynamical generation of scalar masses \footnote{In this paper, the dynamical mass is defined as the renormalized mass for numerically negligible bare mass.} due to different interactions \cite{Bezrukov:2021mio,Bezrukov:2013fka,Mukhanov:2005sc,Lee:2017qve,Bento:2000ah,Bertolami:2016ywc,Munoz:2017ezd,Martin:1997ns,Lee:2017qve} can not be underestimated.
\par
Yukawa interaction vertex is among the phenomenologically interesting vertices \cite{Schwartz:2013pla,Kaku:1993ym,Peskin:1995ev}. Despite that it is mostly considered in fermionic interactions, no unassailable argument exists to prohibit it in scalar sector. A model containing the vertex and the standard model (SM) Higgs field serves as an important avenue to explore the scalar sector. Two Higgs doublet models (2HDM) \cite{Gunion:2002zf,Cabrera:2020lmg,Altmannshofer:2020shb} are among the widely studied models. Relevant to scalar sector, a 2HDM presents two important scenarios once the SM Higgs \cite{Aad:2012tfa,Chatrchyan:2012xdj} is introduced in the theory. First, the model presents a mock up scenario of fermions interacting with the SM Higgs. Second, the model presents an opportunity to investigate how same particles from two different families interact with each other. A highly interesting aspect in both scenarios is that the physics of dark matter in scalar sector arising from possible Higgs-ultralight scalar ($m_{s}=O(10^{-22})$ eV) interactions \cite{Rindler-Daller:2013zxa,Harko:2014vya,Chavanis:2011zi,Huang:2013oqa,Marsh:2015xka,Hui:2016ltb,Primack:2009jr,Hu:2000ke} can also be studied in the model, though it may be more challenging in terms of controlling the involved numerics.
\par
This paper is an extension of Wick Cutkosky (WC) model \cite{Darewych:1998mb,Sauli:2002qa,Efimov:2003hs,Nugaev:2016uqd,Darewych:2009wk} to incorporate two complex doublet fields, termed as the SM Higgs, and the second Higgs fields \footnote{It has the same symmetry as the SM Higgs but different renormalized masses.} for convenience. The two Higgs fields interact with each other only by a real singlet scalar field. The aim is to study the role of Yukawa interaction \footnote{The term is borrowed from the Yukawa interaction of fermions for convenience.} in dynamical scalar mass generation in different regions of the parameter space of the model at different cutoff values, and extract a representative function of the dynamical mass. The renormalized \footnote{The terms renormalized mass and physical mass are interchangeably used throughout the paper for the SM Higgs. The term scalar field is reserved only for the singlet scalar.} masses of the two complex doublet scalars are kept at their physical masses in order to study the phenomenon of DMG from the perspective of phenomenology. The physical mass of the SM Higgs is kept at its experimentally known value \cite{pdg} while the renormalized mass of the second Higgs is chosen for different scenarios mentioned above. A certain advantage of fixing masses is reduction of the parameter space to be explored. It greatly facilitates in training an algorithm \cite{Alpaydin14}, which is no longer a novel approach in quantum field theory \cite{Bachtis:2021xoh,Halverson:2020trp,Akutagawa:2020yeo}, over the samples of calculated dynamical masses in different regions of the parameter space, the details are given in the next section. An advantage is that the quantities subject to training become readily available for richer field theory models which may involve larger Dyson Schwinger equations (DSEs) \cite{Schwinger:1951ex,Schwinger:1951hq,Swanson:2010pw,Roberts:1994dr,Rivers:1987hi} which are more resource hungry.
\par
The approach of DSEs is used for the study. The DSEs for the three field propagators are considered while the interaction vertices are fixed at their tree level form up to certain renormalization dependent terms, the details are given in the next section.
\par
There exist other renormalizable vertices to further extend the model \cite{Schwartz:2013pla,Kaku:1993ym,Peskin:1995ev}. However, such richer models may require further truncations and ansatz \cite{Roberts:1994dr} for the approach of DSEs. As strength of Yukawa vertices to produce dynamical mass is investigated, quartic interactions are not considered at this point.
\section{Technical Details}
The Euclidean version of the Lagrangian \footnote{The implimentation of renormalization procedure differs from commonly used approaches in order to accommodate the intended computations.} with the counter terms is given by
\begin{equation} \label{Lagrangian:eq}
\begin{split}
L = \frac{1}{2}(1+A) \partial_{\mu} \phi(x) \partial^{\mu} \phi(x) + \frac{1}{2} (m_{s}^{2}+B) \phi^{2}(x) + (1+\alpha) \partial_{\mu} h^{\dagger}(x) \partial^{\mu} h(x) \\ + (m_{h}^{2}+\beta) h^{\dagger}(x) h(x) + (1+a) \partial_{\mu} H^{\dagger}(x) \partial^{\mu} H(x) + (m_{H}^{2}+b) H^{\dagger}(x) H(x) \\ + (\lambda_{1}+C_{1}) \phi(x) h^{\dagger}(x) h(x) + (\lambda_{2}+C_{2}) \phi(x) H^{\dagger}(x) H(x)
\end{split}
\end{equation}
where $A$, $B$, $\alpha$, $\beta$, $a$, $b$, $C_{1}$, and $C_{2}$ are coefficients due to the counter terms in the Lagrangian. The real singlet scalar field is represented by $\phi(x)$, $h(x)$ is designated for the SM Higgs boson while $H(x)$ represents the second Higgs boson \footnote{The renormalization procedure is adopted to facilitate the imposed the renormalization conditions.}. The resulting DSEs for the field propagators are given below:
\begin{equation} \label{hdse:eq}
\begin{split}
D_{h}^{-1}(p)=(1+\alpha) p^{2} + m^{2}_{h} (1+\alpha) + 2 (1+A) (1+\alpha) (1+a) \sigma_{h} + \\ (\lambda_{1}+C_{1}) \int_{-\Lambda}^{\Lambda} \frac{d^{4}q}{(2\pi)^{4}} D_{s}(q) \Gamma_{1}(-p,q) D_{h}(q-p)
\end{split}
\end{equation}
\begin{equation} \label{Hdse:eq}
\begin{split}
D_{H}^{-1}(p)=(1+a) p^{2} + m^{2}_{H} (1+a) + 2 (1+A) (1+\alpha) (1+a) \sigma_{H} + \\ (\lambda_{2}+C_{2}) \int_{-\Lambda}^{\Lambda} \frac{d^{4}q}{(2\pi)^{4}} D_{s}(q) \Gamma_{2}(-p,q) D_{H}(q-p)
\end{split}
\end{equation}
\begin{equation} \label{sdse:eq}
\begin{split}
D_{s}^{-1}(p)=(1+A) p^{2} + m^{2}_{s} (1+A) + 2 (1+A) (1+\alpha) (1+a) \sigma_{s} + \\ (\lambda_{1}+C_{1}) \int_{-\Lambda}^{\Lambda} \frac{d^{4}q}{(2\pi)^{4}} D_{h}(q) \Gamma_{1}(q,-p) D_{h}(q-p)+ \\ (\lambda_{2}+C_{2}) \int_{-\Lambda}^{\Lambda} \frac{d^{4}q}{(2\pi)^{4}} D_{H}(q) \Gamma_{2}(q,-p) D_{H}(q-p)
\end{split}
\end{equation}
where the following definitions are used:
\begin{subequations} \label{mterms:eq}
\begin{align}
\beta = \alpha m^{2}_{h} + 2(1+A) (1+\alpha) (1+a) \sigma_{h}  \\
b = a m^{2}_{H} + 2(1+A) (1+\alpha) (1+a) \sigma_{H}  \\
B = A m^{2}_{s} + 2(1+A) (1+\alpha) (1+a) \sigma_{s}
\end{align}
\end{subequations} 
$\sigma_{h}$, $\sigma_{H}$, $\sigma_{s}$ are the terms to be determined during a computation. Due to their nature, above definitions do not impose any additional constraints on the equations. The definition of the two vertices during computations is given below:
\begin{subequations} \label{vers:eq}
\begin{align}
\Gamma_{1}(u,v)=(1+A) (1+\alpha) (1+a) \tilde{\Gamma}_{1}(u,v) \\
\Gamma_{2}(u,v)=(1+A) (1+\alpha) (1+a) \tilde{\Gamma}_{2}(u,v)
\end{align}
\end{subequations}
Hence, the DSEs for the three field propagators become
\begin{equation} \label{hfdse:eq}
\begin{split}
D^{-1}_{h}(p)=(1+\alpha) [\ p^{2} + \frac{m^{2}_{h,r}}{(1+\alpha)} + (\lambda_{1}+C_{1})(1+A)(1+a) \\ \int_{-\Lambda}^{\Lambda} \frac{d^{4}q}{(2\pi)^{4}} D_{s}(q) \tilde{\Gamma}_{1}(-p,q) D_{h}(q-p) ]\
\end{split}
\end{equation}
\begin{equation} \label{Hfdse:eq}
\begin{split}
D^{-1}_{H}(p)=(1+a) [\ p^{2} + \frac{m^{2}_{H,r}}{1+a} + (\lambda_{2}+C_{2})(1+A)(1+\alpha) \\ \int_{-\Lambda}^{\Lambda} \frac{d^{4}q}{(2\pi)^{4}} D_{s}(q) \tilde{\Gamma}_{2}(-p,q) D_{H}(q-p) ]\
\end{split}
\end{equation}
\begin{equation} \label{sfdse:eq}
\begin{split}
D^{-1}_{s}(p)=(1+A) [\ p^{2} + m^{2}_{s} + 2 (1+a) (1+\alpha) \sigma_{s} + (\lambda_{1}+C_{1})(1+a)(1+\alpha) \\ \int_{-\Lambda}^{\Lambda} \frac{d^{4}q}{(2\pi)^{4}} D_{h}(q) \tilde{\Gamma}_{1}(q,-p) D_{h}(q-p) + (\lambda_{2}+C_{2})(1+a)(1+\alpha) \\ \int_{-\Lambda}^{\Lambda} \frac{d^{4}q}{(2\pi)^{4}} D_{H}(q) \tilde{\Gamma}_{2}(q,-p) D_{H}(q-p) ]\
\end{split}
\end{equation}
where in equation \ref{hfdse:eq} the renormalized mass for the SM Higgs ($ m_{h,r} $) is fixed at 125.09 GeV during the entire study, while the renormalized mass of the second Higgs boson is fixed during each computation \footnote{The definition of the squared physical mass of the SM Higgs is $m^{2}_{h,r}=m^{2}_{h}+\beta$, and the definition of the squared renormalized mass of the second Higgs is $m^{2}_{H,r}=m^{2}_{H}+b$.}. Equations \ref{hfdse:eq}-\ref{sfdse:eq} are the three DSEs considered for the study. The quantities $\tilde{\Gamma}_{1}(u,v)$ and $\tilde{\Gamma}_{2}(u,v)$ are fixed at $\lambda_{1}$ and $\lambda_{2}$, respectively \cite{Roberts:1994dr}. However, in the current investigation the vertices can still change depending upon the contributions from the coefficients in the counter terms, see equations \ref{vers:eq}.
\par
In our investigation, the dynamical renormalized (squared) scalar mass is defined as the renormalized (squared) mass with insignificant contribution from the corresponding bare mass value. The definition and the method of counter terms involving scalar mass are not in tension. It has been observed that a numerically insignificant value of the scalar bare mass does not influence any outcome. Hence, setting $m_{s}$ to 0 is found equivalent for computations. During the computations, $m_{s}=10^{-8}$ GeV is used.
\par
For each of the propagators, the following renormalization conditions are used.
\begin{equation} \label{hcond:eq}
D_{h}^{ij}(p)  |_{p^{2}=m^{2}_{h,r}} = \frac{\delta ^{ij}}{p^{2}+m^{2}_{h,r}} |_{p^{2}=m^{2}_{h,r}}
\end{equation}
\begin{equation} \label{Hcond:eq}
D_{H}^{ij}(p)  |_{p^{2}=m^{2}_{H,r}} = \frac{\delta ^{ij}}{p^{2}+m^{2}_{H,r}} |_{p^{2}=m^{2}_{H,r}}
\end{equation}
\begin{equation} \label{scond:eq}
D_{s}(p)  |_{p=1} = \frac{1}{p^{2}} |_{p=1}
\end{equation}
The following two conditions are aimed to extract the correlation functions and the other quantities which are introduced for the counter terms.
\begin{equation} \label{hleast:eq}
\begin{split}
\int_{-\Lambda}^{\Lambda} (\ -D^{-1}_{h}(p)+ (1+\alpha) [\ p^{2} + \frac{m^{2}_{h,r}}{(1+\alpha)} + (\lambda_{1}+C_{1})(1+A)(1+a) \\ \int_{-\Lambda}^{\Lambda} \frac{d^{4}q}{(2\pi)^{4}} D_{s}(q) \tilde{\Gamma}_{1}(-p,q) D_{h}(q-p) ]\ )\ ^{2} dp =0
\end{split}
\end{equation}
\begin{equation} \label{Hleast:eq}
\begin{split}
\int_{-\Lambda}^{\Lambda} (\ -D^{-1}_{H}(p)+(1+a) [\ p^{2} + \frac{m^{2}_{H,r}}{1+a} + (\lambda_{2}+ C_{2})(1+A)(1+\alpha) \\ \int_{-\Lambda}^{\Lambda} \frac{d^{4}q}{(2\pi)^{4}} D_{s}(q) \tilde{\Gamma}_{2}(-p,q) D_{H}(q-p) ]\ )\ ^{2} dp =0
\end{split}
\end{equation}
Equations \ref{hleast:eq} - \ref{Hleast:eq} are indeed the implementation of the least square method with errors $E_{1}$ and $E_{2}$ defined below.
\begin{equation} \label{herr:eq}
\begin{split}
E_{1}= \int_{-\Lambda}^{\Lambda} (\ -D^{-1}_{h}(p)+ (1+\alpha) [\ p^{2} + \frac{m^{2}_{h,r}}{(1+\alpha)} + (\lambda_{1}+C_{1})(1+A)(1+a) \\ \int_{-\Lambda}^{\Lambda} \frac{d^{4}q}{(2\pi)^{4}} D_{s}(q) \tilde{\Gamma}_{1}(-p,q) D_{h}(q-p) ]\ )\ ^{2} dp
\end{split}
\end{equation}
\begin{equation} \label{Herr:eq}
\begin{split}
E_{2}=\int_{-\Lambda}^{\Lambda} (\ -D^{-1}_{H}(p)+(1+a) [\ p^{2} + \frac{m^{2}_{H,r}}{1+a} + (\lambda_{2}+C_{2})(1+A)(1+\alpha) \\ \int_{-\Lambda}^{\Lambda} \frac{d^{4}q}{(2\pi)^{4}} D_{s}(q) \tilde{\Gamma}_{2}(-p,q) D_{H}(q-p) ]\ )\ ^{2} dp
\end{split}
\end{equation}
With imposition of these constraints, the problem at hand becomes that of optimization in which solutions are sought which satisfy equations \ref{hleast:eq}-\ref{Hleast:eq}.
\par
An additional condition given below is also imposed in order to ensure positivity of the renormalized squared dynamical scalar mass and evade unwanted numerical fluctuations which may arise due to the difference between the fixed renormalized masses and the dynamically generated masses, and the ever-present limitation in momentum resolution.
\begin{equation} \label{mass2cond:eq}
\begin{split}
m^{2}_{s,r} = (1 + A) (\ m^{2}_{s} + 2(1+\alpha) (1+a) \sigma_{s} )\ \geq 0
\end{split}
\end{equation}
\par
In order to further suppress numerical fluctuations, the SM Higgs is expanded in the form given below:
\begin{equation} \label{hexp1:eq}
D^{ij}_{h}(p)= \delta^{ij} \frac{1}{c(p^{2}+d+f(p))}
\end{equation}
with $f(p)$ given by
\begin{equation} \label{hexp2:eq}
f(p) = \frac{\displaystyle \sum_{l=0}^{N} a_{l} p^{2l}}{\displaystyle \sum_{l=0}^{N} b_{l} p^{2l}}
\end{equation}
In equations \ref{hexp1:eq}-\ref{hexp2:eq}, $c$, $d$, $a_{l}$, and $b_{l}$ are the coefficients to be determined during a computation. A similar expansion with different coefficients is used for the second Higgs propagator. Beside stability, these expansions are also time efficient while performing renormalization and updating the SM and the second Higgs propagators.
\par
The computation starts with $\sigma_{H}=\sigma_{s}=C_{1}=C_{2}=0$, i.e. with no contribution by the counter terms to the renormalized masses and the two Yukawa couplings. Both Higgs propagators are also rendered their respective tree level structures. For the SM Higgs in equations \ref{hexp1:eq}-\ref{hexp2:eq}, $c=1$ and $d=m^{2}_{h}$ \footnote{$m^{2}_{h}=m^{2}_{h,r}$ is used throughout the study.} while all the other coefficients are zero. A similar setup of coefficients is used for the second Higgs with $d=m^{2}_{H}$. The terms $1+\alpha$ and $1+a$ are calculated from the renormalization conditions in equations \ref{hcond:eq} and \ref{Hcond:eq}. The scalar propagator assumes the values using the equation \ref{sdse:eq} and the quantity $1+A$ is calculated by the renormalization condition \ref{scond:eq} \footnote{The scalar propagator is calculated without the term $1+A$ and then then renormalization condition sets the value of the term $1+A$.}.
\par
An iteration involves updating of the correlation functions and parameters. During an iteration, first, the $\sigma_{s}$, $C_{1}$, and $C_{2}$ are updated as in the mentioned ordered. The update of each of these quantities is performed using Newton Raphson's method with the criteria imposed by the least square method in equations \ref{hleast:eq}-\ref{Hleast:eq}. The updated value is accepted only when both of the errors $E_{1}$ and $E_{2}$ reduce, see equations \ref{herr:eq}-\ref{Herr:eq}.
\par
It is followed by the SM Higgs propagator for which the coefficients in equations \ref{hexp1:eq}-\ref{hexp2:eq} are updated with the above mentioned criteria of acceptance. Upon each change, the SM Higgs propagator is calculated from equations \ref{hexp1:eq}-\ref{hexp2:eq} and renormalized using equation \ref{hcond:eq}.
\par
Lastly, the second Higgs propagator is updated using the same procedure as described above for the SM Higgs, but using equation \ref{Hcond:eq} for renormalization. Upon every change, the scalar propagator is calculated from equation \ref{sfdse:eq} and renormalized using equation \ref{scond:eq}, as mentioned earlier.
\par
A computation concludes only when there are either no further improvements in the quantities such that $E_{1}$ and $E_{2}$ are further reduced, or both of these errors are equal or below the preset value of tolerance. The value of tolerance is set at $10^{-20}$.
\par
Gauss quadrature algorithm is used for numerical integration in the DSEs. The algorithms are developed in C++ environment.
\par
The solutions presented are unique in the sense that order of updates performed on the propagators and other quantities practically does not effect the quantities being computed.
\par
As the Lagrangian of the model does not contain quartic self interactions it is assumed that the model is a non-trivial theory \cite{Hasenfratz:1988kr,Gliozzi:1997ve,Weber:2000dp,Jora:2015yga,Aizenman:1981zz,Weisz:2010xx,Siefert:2014ela,Hogervorst:2011zw}.
\par
The assumption that the considered model may be a part of a larger new scalar sector, containing further interaction vertices, is justifiable. As other vertices may also contribute in physical processes, the effects of the Yukawa vertices rendering the complete theory Hamiltonian nagative are expected to be at least significantly mitigated even before a full theory containing the SM is considered. Hence, from the perspective of path integral approach \cite{Rivers:1987hi}, the questionable feature of negative Hamiltonian for certain paths is entirely neglected at this point. It is important to note that the large magnitude of the (squared) Higgs mass may already contribute in mitigating the effects.
\par
Once the scalar dynamical masses in the parameter space are calculated, knowledge from machine learning \cite{Alpaydin14} is employed to represent the results in an attempt to understand how scalar dynamical masses manifest in the parameter space and either estimate or impose bounds on the critical coupling value. An obvious advantage is the availability of the quantities for richer field theories which mostly require truncations and ansatz in DSE related studies. There are three free parameters in the model. Hence, scalar dynamical mass is expanded in terms of the second Higgs mass $m_{H}$, and the two Yukawa couplings ($\lambda_{1}$ and $\lambda_{2}$) as given below:
\begin{equation} \label{sDMGexpansion:eq}
m_{s,f} (m_{H},\lambda_{1},\lambda_{2}) = \sum_{i=0}^{i=6} \sum_{j=0}^{j=6} \sum_{k=0}^{k=6} a_{ijk} m_{H}^{i} \lambda_{1}^{j} \lambda_{2}^{k}
\end{equation}
where $m_{s,f}$ is the function representing scalar dynamical mass in the parameter space. The error value $E_{f}$ is described by the following expression:
\begin{equation} \label{sdmgerror:eq}
E_{f} =\frac{1}{N}\sqrt{\sum_{i=1}^{i=N}(m_{s,r} - m_{s,f})_{i}^{2}}
\end{equation}
where $N$ is the number of explored points in the parameter space and $i$ represents the $i^{th}$ point in the parameter space. The model is studied on 64 points for each of the cutoff values set at $10$ TeV and $100$ TeV. Due to the combination of cutoff effects and numerical artifacts, (supervised) machine learning \cite{Alpaydin14} is performed separately at the two cutoff values.
\par
The training proceeds as follows: The starting value of all weights $a_{ijk}$ are set to zero which results in non-zero cost function $E_{f}$ due to non-vanishing $m_{s,r}$. Against each value of $i$, $j$, and $k$, the weights are systematically updated using Newton Raphson method \footnote{It was found that other methods, such as bisection method, leads to numerically similar results.} and a new value of $E_{f}$ is computed against the update of each $a_{ijk}$. The update is accepted if $E_{f}$ decreases for a combination of $i$, $j$, and $k$. If it does not, the update is reverted in favor of the previous value of $a_{ijk}$, and the weight for another combination of $i$, $j$, and $k$ is considered. An iteration is complete once all intended combinations of $i$, $j$, and $k$ are exhausted. The tolerance is set at the value $10^{-16}$. The computation is concluded once the improvement of $E_{f}$ is either smaller than the tolerance value, or $E_{f} \leq 10^{-16}$ is reached. Throughout the training, $0 \leq m_{s,f} (m_{H},\lambda_{1},\lambda_{2}) $ is imposed. In order to remove any personal bias, no value of scalar dynamical mass was taken as outliers.
\par
Once the training is accomplished, estimates on the critical coupling value are made.
\section{Field Propagators}
\begin{figure}
\centering
\includegraphics[width=\linewidth]{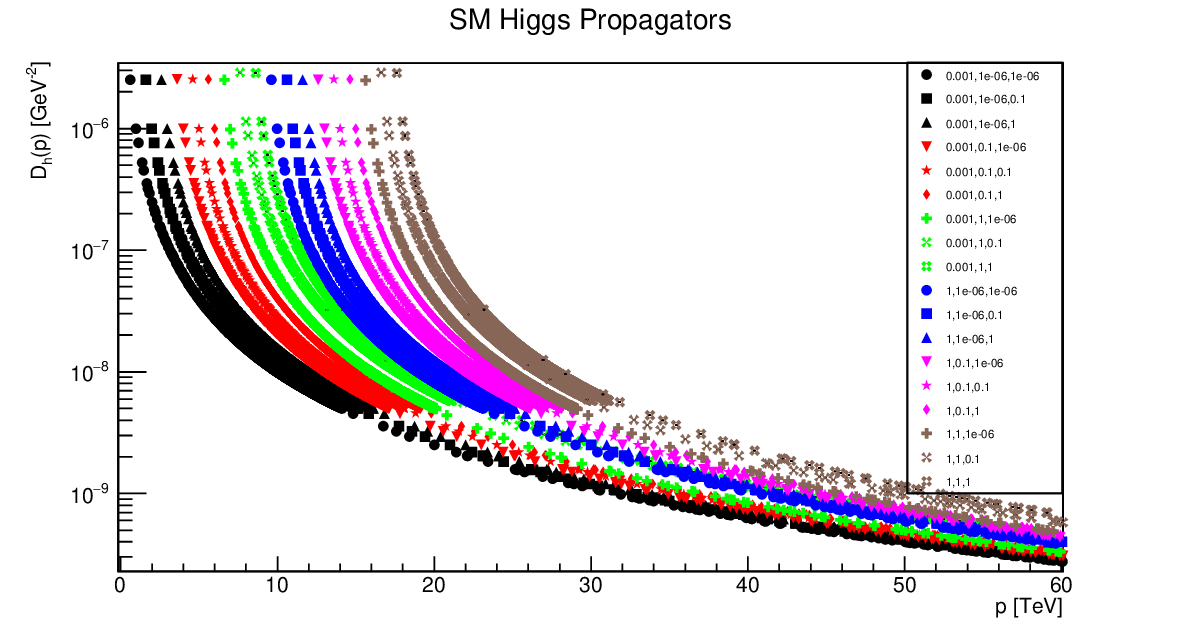}
\caption{\label{fig:h1prs1} The SM Higgs propagators with $m_{H}=0.001$ GeV and $m_{H}=1.0$ GeV are plotted for cutoff values at $10$ TeV and $100$ TeV. The parameters in the legend are given as $(m_{H},\lambda_{1},\lambda_{2})$ with all of the parameters mentioned in GeV. For the same cutoff, every two consecutive propagators are $1.0$ TeV apart on the momentum axis for the sake of clarity.}
\end{figure}
\begin{figure}
\centering
\includegraphics[width=\linewidth]{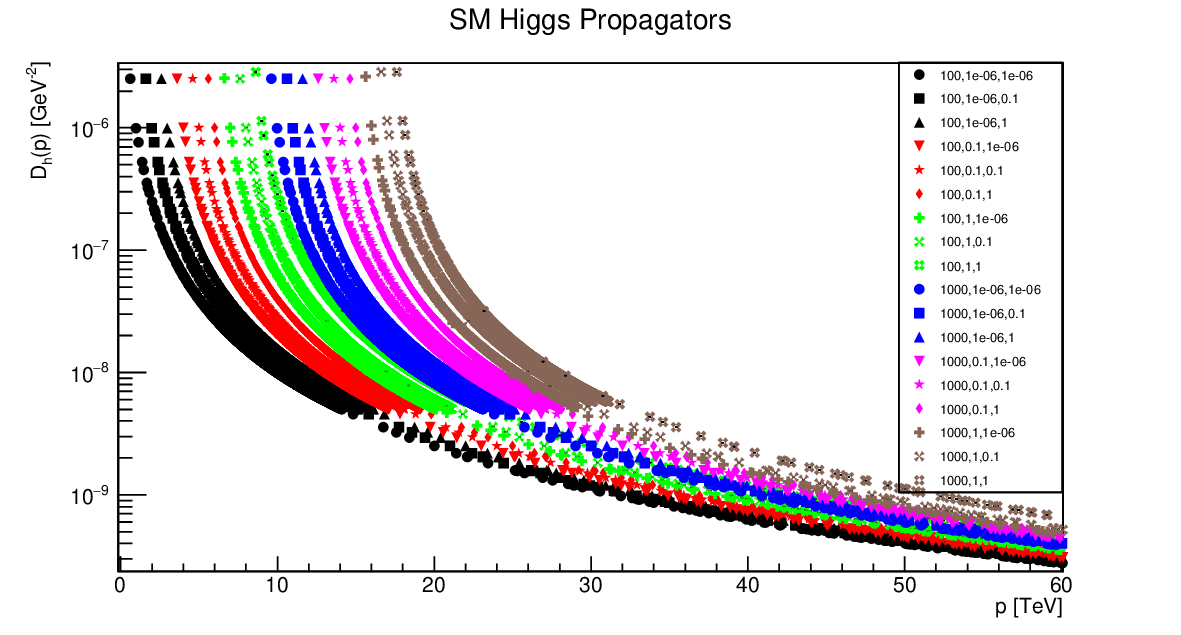}
\caption{\label{fig:h1prs2} The SM Higgs propagators with $m_{H}=100$ GeV and $m_{H}=1000$ GeV are plotted for cutoff values at $10$ TeV and $100$ TeV. The parameters in the legend are given as $(m_{H},\lambda_{1},\lambda_{2})$ with all of the parameters mentioned in GeV. For the same cutoff, every two consecutive propagators are $1.0$ TeV apart on the momentum axis for the sake of clarity.}
\end{figure}
\begin{figure}
\centering
\includegraphics[width=\linewidth]{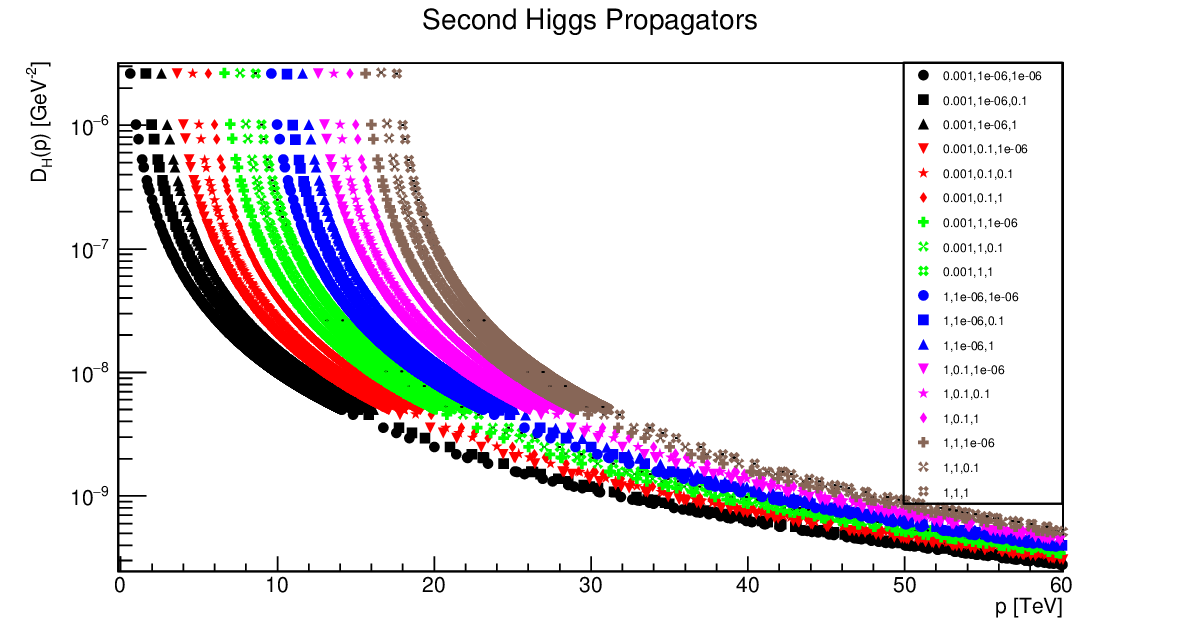}
\caption{\label{fig:h2prs1} The second Higgs propagators with $m_{H}=0.001$ GeV and $m_{H}=1.0$ GeV are plotted for cutoff values at $10$ TeV and $100$ TeV. The parameters in the legend are given as $(m_{H},\lambda_{1},\lambda_{2})$ with all of the parameters mentioned in GeV. For the same cutoff, every two consecutive propagators are $1.0$ TeV apart on the momentum axis for the sake of clarity.}
\end{figure}
\begin{figure}
\centering
\includegraphics[width=\linewidth]{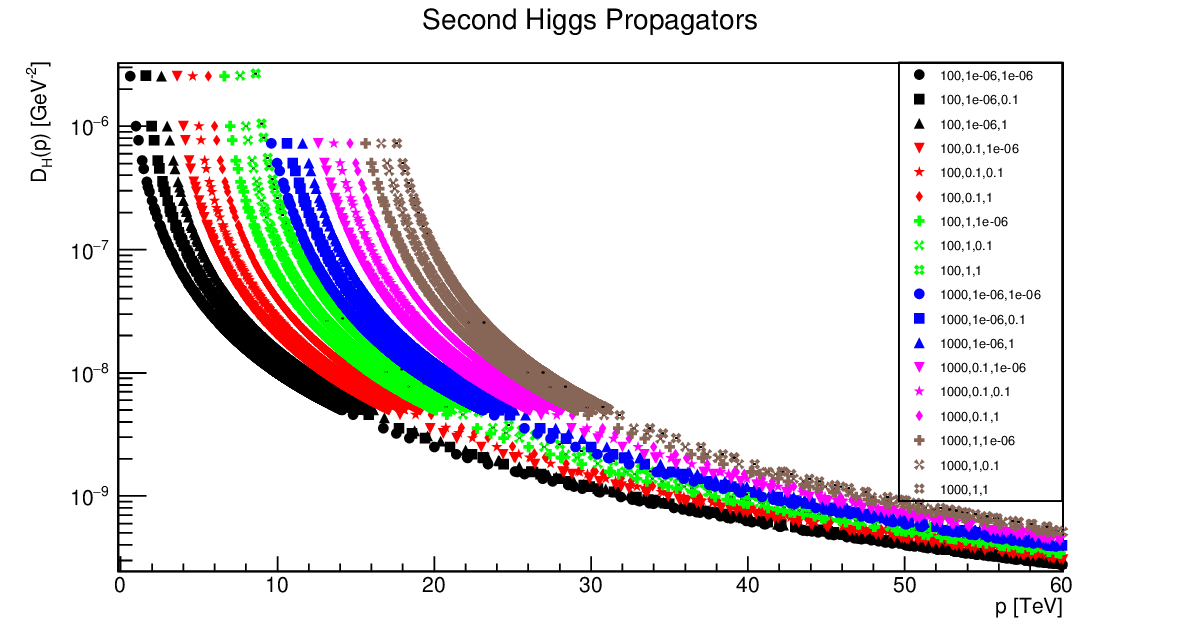}
\caption{\label{fig:h2prs2} The second Higgs propagators with $m_{H}=100$ GeV and $m_{H}=1000$ GeV are plotted for cutoff values at $10$ TeV and $100$ TeV. The parameters in the legend are given as $(m_{H},\lambda_{1},\lambda_{2})$ with all of the parameters mentioned in GeV. For the same cutoff, every two consecutive propagators are $1.0$ TeV apart on the momentum axis for the sake of clarity.}
\end{figure}
\begin{figure}
\centering
\includegraphics[width=\linewidth]{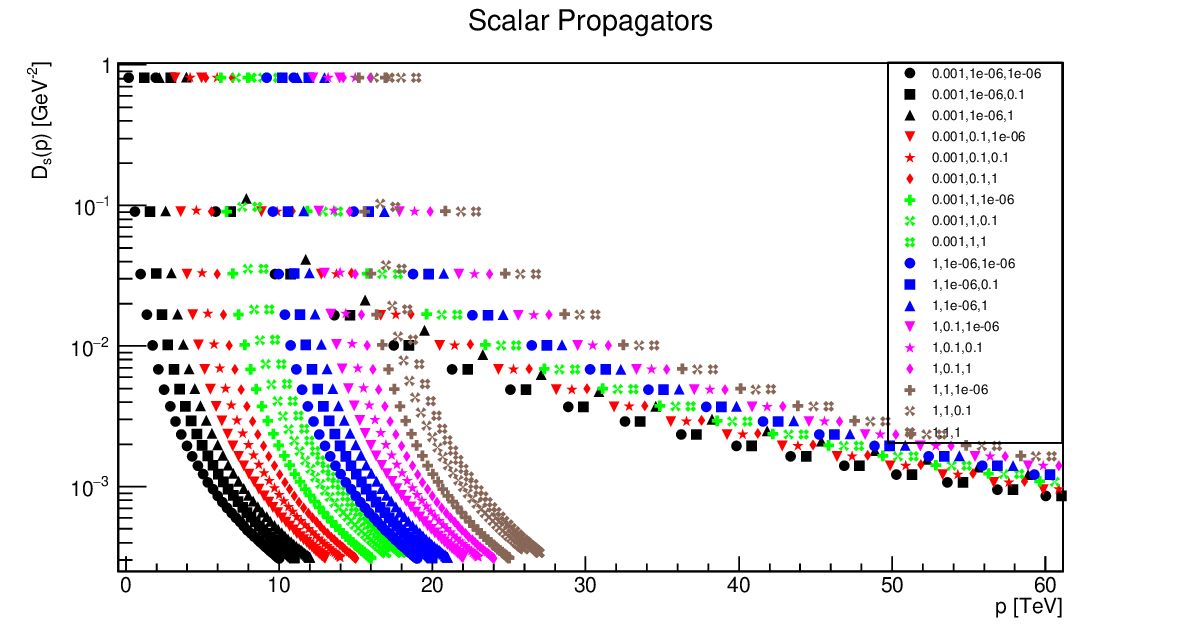}
\caption{\label{fig:sprs1} The scalar propagators with $m_{H}=0.001$ GeV and $m_{H}=1.0$ GeV are plotted for cutoff values at $10$ TeV and $100$ TeV. The parameters in the legend are given as $(m_{H},\lambda_{1},\lambda_{2})$ with all of the parameters mentioned in GeV. For the same cutoff, every two consecutive propagators are $1.0$ TeV apart on the momentum axis for the sake of clarity.}
\end{figure}
\begin{figure}
\centering
\includegraphics[width=\linewidth]{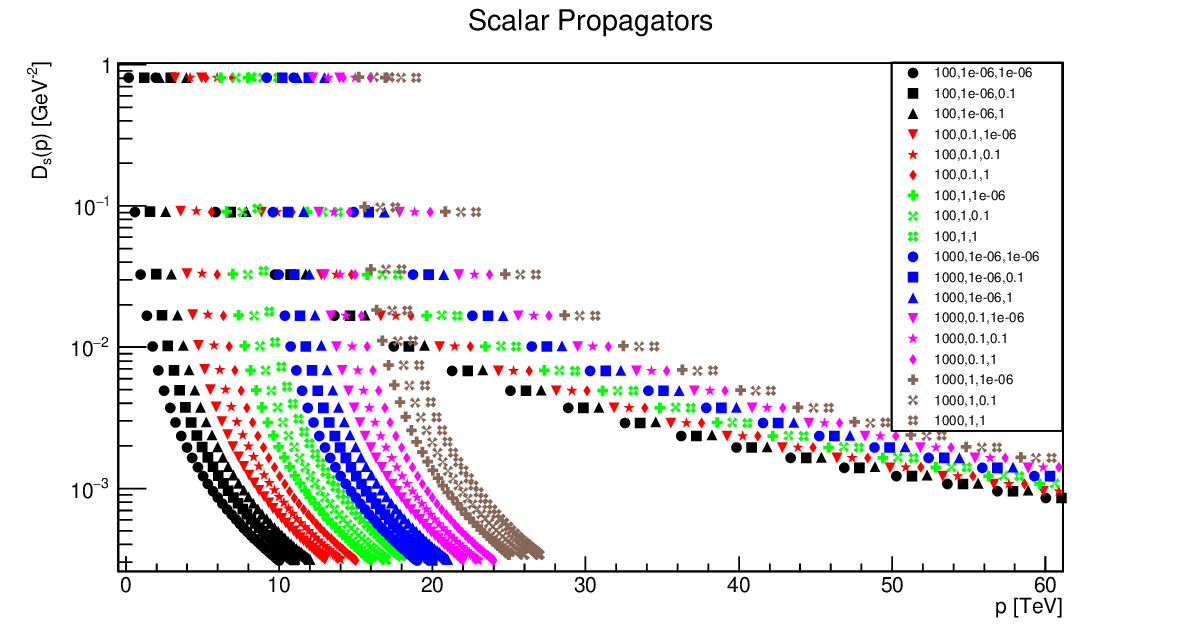}
\caption{\label{fig:sprs2} The scalar propagators with $m_{H}=100$ GeV and $m_{H}=1000$ GeV are plotted for cutoff values at $10$ TeV and $100$ TeV. The parameters in the legend are given as $(m_{H},\lambda_{1},\lambda_{2})$ with all of the parameters mentioned in GeV. For the same cutoff, every two consecutive propagators are $1.0$ TeV apart on the momentum axis for the sake of clarity.}
\end{figure}
Unlike the modelled vertices, the field propagators receive momentum dependent contributions due to Yukawa interactions while conforming to the renormalization conditions in the model \footnote{The vertex may depend upon the parameters (couplings and bare mass) at most, see equation \ref{vers:eq}.}. These contributions are relatively more diverse for scalar propagators as they receive contributions from both Higgs propagators due to the two interactions. However, it may also introduce numerical fluctuations induced by the two Higgs propagator DSEs as well as other involved quantities in the model.
\par
The cutoff effects are not significant in the SM Higgs propagators throughout the explored parameter space, see figures \ref{fig:h1prs1}-\ref{fig:h1prs2}, particularly for $\lambda_{1} < 1$. The deviations are mostly due to the multiplicative constant which is computed using the renormalization condition \ref{hcond:eq}. In the parameter space with $\lambda_{1}$ as low as $10^{-6}$, the propagator is least sensitive. As the couplings are increased, slight enhancement is observed while considerable changes occur only at $\lambda_{1}=1.0$. The propagators are found to have similar qualitative behavior for all the renormalized masses of the second Higgs which indicates that the second Higgs field does not influence the SM propagators which could have been possible through the DSE of the scalar propagator. Overall, there is no significant changes in the SM Higgs propagators in the parameter space of the model.
\par
The second Higgs field differs from the SM Higgs field due to different renormalized masses and couplings. Hence, particularly for masses in the vicinity of the SM Higgs mass, it is expected to have similar behavior within numerical fluctuations. The propagators are shown in figures \ref{fig:h2prs1}-\ref{fig:h2prs2}. There are no considerable cutoff effects, as is the case with the SM Higgs propagators. However, a certain dependence on $m_{H}$ is observed in the propagators which is relatively stronger for $m_{H} < m_{h}$ and weakens as $m_{H}$ approaches $m_{h}$. The two Higgs propagators have similar behavior for $m_{H} \simeq m_{h}$ which also verifies implementation of the algorithms. For $m_{H} = 1$ TeV  dependence on couplings is lost since the bare mass of the second Higgs field dominates in contribution to the propagators rendering it practically a tree level structure up to the renormalization dependent term. The propagators are relatively suppressed in the low momentum region for larger bare mass which is expected for the case of dominant tree level contribution to the propagator.
\par
The scalar singlet propagators are shown in figures \ref{fig:sprs1}-\ref{fig:sprs2}. As argued above, cutoff effects are evident on the propagators \footnote{There may also be contributions from the numerical interpolation performed during renormalization of the scalar propagators. In order to suppress these effects, resolution in momentum is kept at its highest, given the available resources.}. The overall behavior is that for higher couplings, particularly $\lambda_{2}$ since $m_{h}$ is kept fixed, the propagators are enhanced depending upon the second Higgs mass $m_{H}$. As $m_{H}$ increases, this effect tends to disappear in favor of a tree level structure, as is the case for the other field propagators. The scalar propagators suffer strongly from the cutoff effects despite that both Higgs propagators are not effected to such an extent. It implies that the cutoff effects must show up in at least one of the calculated quantities other than the two Higgs propagators. The field propagators are found monotonically decreasing and without containing any peculiarities such as zero crossing.
\section{Dynamical Scalar Masses}
\begin{figure}
\centering
\includegraphics[width=\linewidth]{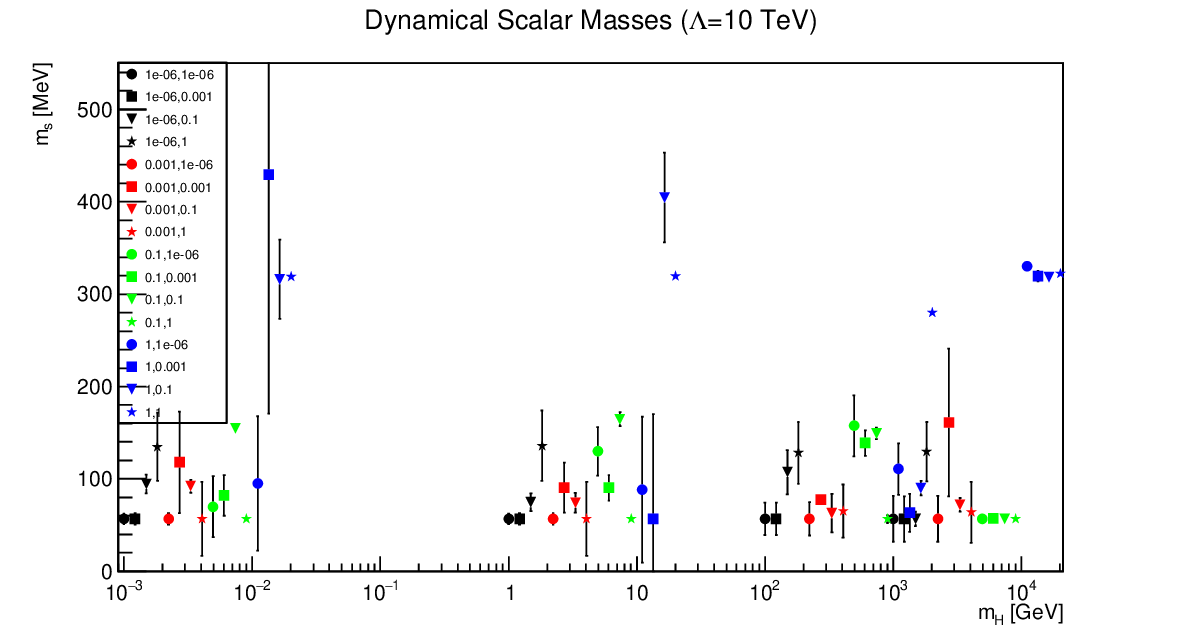}
\caption{\label{fig:smasses10TeV} Dynamical scalar masses are plotted against second Higgs mass $m_{H}$ for various couplings at $10$ TeV cutoff. The couplings (in GeV) are shown in the caption as $\lambda_{1},\lambda_{2}$. Every consecutive pair of couplings is slightly displaced along x-axis for clarity. The error bars represent difference between the values of scalar dynamical masses obtained by computation (plotted values) and by training of the algorithms.}
\end{figure}
\begin{figure}
\centering
\includegraphics[width=\linewidth]{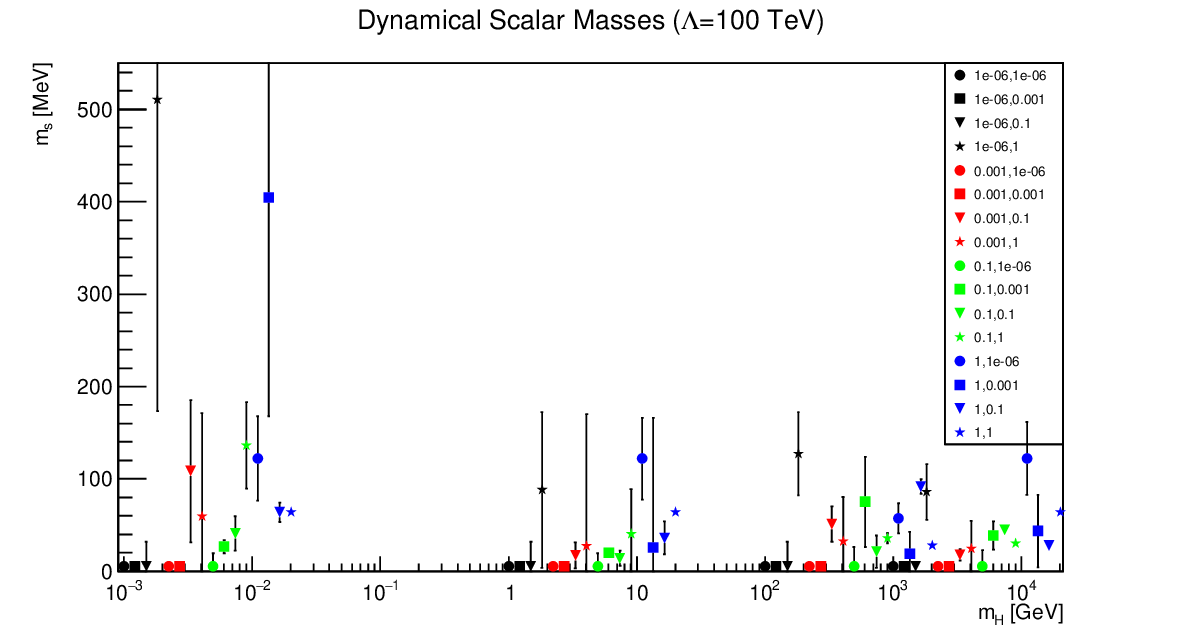}
\caption{\label{fig:smasses100TeV} Dynamical scalar masses are plotted against second Higgs mass $m_{H}$ for various couplings at $100$ TeV cutoff. The couplings (in GeV) are shown in the caption as $\lambda_{1},\lambda_{2}$. Every consecutive pair of couplings is slightly displaced along x-axis for clarity. The error bars represent difference between the values of scalar dynamical masses obtained by computation (plotted values) and by training of the algorithms.}
\end{figure}
\begin{figure}
\centering
\includegraphics[width=\linewidth]{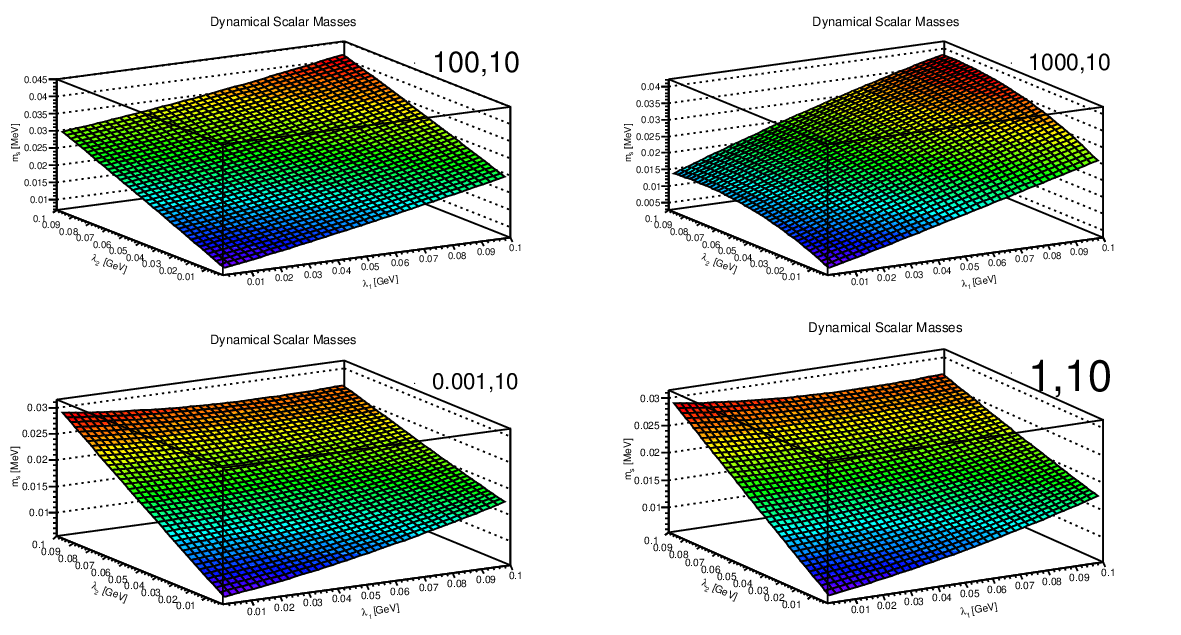}
\caption{\label{fig:smassfit10TeV} Dynamical scalar masses obtained by training algorithms are shown against the two couplings $\lambda_{1}$ and $\lambda_{2}$, for various second Higgs masses $m_{H}$ in GeV and at cutoff $\Lambda=10$ TeV indicated on the caption.}
\end{figure}
\begin{figure}
\centering
\includegraphics[width=\linewidth]{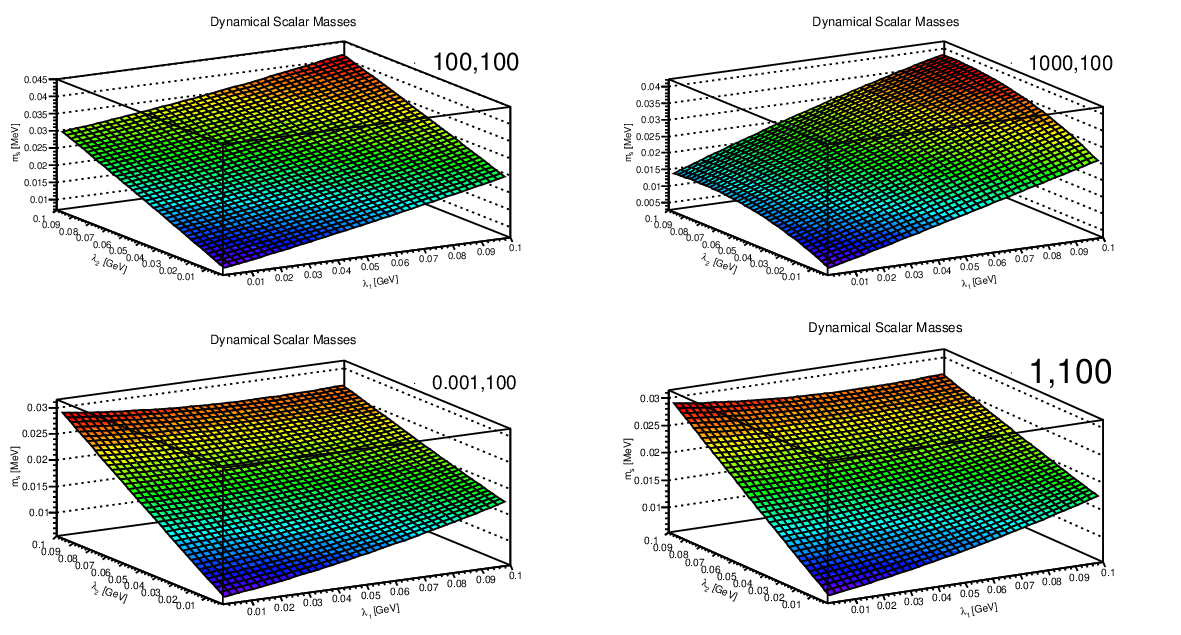}
\caption{\label{fig:smassfit100TeV} Dynamical scalar masses obtained by training algorithms are shown against the two couplings $\lambda_{1}$ and $\lambda_{2}$, for various second Higgs masses $m_{H}$ in GeV and at cutoff $\Lambda=100$ TeV indicated on the caption.}
\end{figure}
Since the scalar propagator also contains its dynamically generated mass, the arguments related to the cutoff effects and numerical artifacts translate into the dynamical masses. However, there are certain vivid features in the masses shown in the figures \ref{fig:smasses10TeV}-\ref{fig:smasses100TeV}.
\par
Considering the dynamical masses for 100 TeV cutoff in figure \ref{fig:smasses100TeV}, the extent of the mass production is relatively higher for the second Higgs mass in MeVs and high second coupling for a number of cases. For very small couplings, large difference between the computed value and $m_{s,f}$, possibly due to numerical fluctuations, suggests that these two points could have been taken as outliers. Hence, it is concluded with confidence that the dynamically generated scalar mass in the model is restricted well within 200 MeVs. It is clear that the Yukawa interaction in the scalar sector has a limitation when it comes to producing dynamical masses in GeVs.
\par
It is also evident from figures \ref{fig:smasses10TeV}-\ref{fig:smasses100TeV} that cutoff effects are weaker for the cutoff in hundreds of TeVs. For higher cutoff, the model produces scalar masses with less dependence on the second Higgs mass. Furthermore, the production of scalar mass is significantly lower if one of the couplings is as low as $10^{-6}$ GeV, unless the other coupling reaches $1.0$ GeV, see figure \ref{fig:smasses100TeV}. Hence, the model does posses a demarcation over the coupling values in the vicinity of $1.0$  and $10^{-6}$ GeV. For the case of both couplings at $10^{-6}$ GeV, scalar mass practically looses dependence on the two couplings. It is taken as a sign of existence of critical coupling value between $10^{-3}$ GeV and $10^{-6}$ GeV.
\par
An interesting feature in the low coupling region is that for higher cutoff values the model produces smaller masses which also tend to decrease with the two couplings. One may expect that the behavior persists until the critical coupling value. If this is indeed the case, the model may be useful to study ultralight scalar interactions \cite{Rindler-Daller:2013zxa,Harko:2014vya,Chavanis:2011zi,Huang:2013oqa,Marsh:2015xka,Hui:2016ltb,Primack:2009jr,Hu:2000ke}, though it may undoubtedly be a daunting task from the perspective of numerical precision.
\par
Scalar dynamical masses obtained by a trained algorithm are plotted in figures \ref{fig:smassfit10TeV}-\ref{fig:smassfit100TeV}. The weights of the expansion in equation \ref{sDMGexpansion:eq} are given in appendix A. An immediate observation is an indication of a unique value of the critical coupling in the model below $10^{-3}$ GeV. Determining exact value is hampered by the ever-present limitation associated with data and precision in machine learning. However, it is evident that the model strongly favors a critical coupling in the region $10^{-6} < \lambda_{i} < 10^{-3}$ GeV.
\par
However, the weights $a_{ijk}$ are found to have a particular behavior. Firstly there are strong contributions from the terms with $k = 0$ and $i,j \neq 0$, while there is practically no contributions from $a_{ijk}$ for $i,j=0$ irrespective of the value of $k$. Increasing the cutoff suppresses $a_{ijk}$ for most of the $i$ and $k$ values at $j=0$. Since the model has significant cutoff effects which become milder as the cutoff is raised, it is expected that scalar dynamical mass may not receive significant contributions from the terms involving $j=0$ at high cutoff values.
\section{Conclusion}
In the presence of an SM Higgs, the two Higgs doublet model has the capacity to dynamically produce scalar masses in the vicinity of, or even larger than, that of the lightest leptons or quarks. As the cutoff is raised above $100$ TeV, stability in the mass with magnitude below $200$ MeV ensues against the second Higgs mass. However, the dynamical mass has sensitivity to the couplings which diminishes as the couplings are reduced below $10^{-3}$ GeV, with the minimum mass being produced. It strongly points towards existence of a critical coupling between $10^{-3}$ GeV and $10^{-6}$ GeV. It presents an opportunity to investigate the model to understand new physics involving the particles considerably lighter than $1$ GeV, such as ultra-light scalars. At the same time, it invites for further study of richer models which contain higher renormalizable vertices in an attempt to explore existence of critical couplings and extent of mass generation in the presence of variety of interactions.
\par
The role of cutoff effects can not be neglected in the model. Scalar propagator and scalar dynamical mass suffer the most, while for (both) Higgs masses larger than $100$ GeV the field propagators are relatively less effected as was observed in the two Higgs propagators. Since dynamical masses are less than $1$ GeV, numerical fluctuations and the cutoff effects hamper in finding an accurate description of how the masses behave in the parameter space. However, considering sufficient number of points in the parameter space a mathematical description for dynamical masses was still found which concurs with the deduction regarding the critical coupling in the model.
\par
A model with the capacity to dynamically produce masses considerably larger than the lightest quarks and leptons can certainly not be ruled out in scalar interactions. The study invites extensions, such as by including richer interactions or other fields, to have a better understanding of how (particularly light) scalars play role at the fundamental level in our universe.
\section{Acknowledgments}
This work was supported by Lahore University of Management Sciences, Pakistan for developing the algorithms and performing computations.
\section{Appendix A} The values of the weights $a_{ijk}$ in equation \ref{sDMGexpansion:eq} for cutoff values at $\Lambda=10$ TeV and $\Lambda=100$ TeV are given below:
\begin{center}
\begin{longtable}{ | m{0.5cm} | m{0.5cm}| m{0.5cm} | m{3.5cm} | m{3.5cm} | }
\hline
$i$ & $j$ & $k$ & $a_{ijk}$ (10 TeV) & $a_{ijk}$ (100   TeV) \\
\hline
0 & 0 & 0 & 0.062470119000000 & -0.005102152199999 \\
\hline
1 & 0 & 0 & 0.432639286999998 & 0.067154687499999 \\
\hline
0 & 1 & 0 & 0.431424060300000 & 0.276498992400000\\
\hline
0 & 0 & 1 & 0.000127544300000 & 0.000023271900000\\
\hline
2 & 0 & 0 & -0.196267245799999 & 0.826964182800007\\
\hline
1 & 1 & 0 & 3.483278381999988 & -2.646594633199990\\
\hline
1 & 0 & 1 & 0.001256076799999 & 0.000786048000000\\
\hline
0 & 2 & 0 & -2.426385300799987 & -0.005208820599999\\
\hline
0 & 1 & 1 & -0.001780857000000 & -0.000014844700000\\
\hline
0 & 0 & 2 & -0.0000001083 & -0.000000023100000\\
\hline
3 & 0 & 0 & -0.595301560000001 & -0.158007649599999\\
\hline
2 & 1 & 0 & 0.311572103699999 & 5.428065021000011\\
\hline
2 & 0 & 1 & -0.001414899099999 & -0.002876003599999\\
\hline
1 & 2 & 0 & -2.967471651800008 & -3.556474306000007\\
\hline
1 & 1 & 1 & -0.011901846299999 & 0.013548386599999\\
\hline
1 & 0 & 2 & -0.0000018441 & -0.000000603000000\\
\hline
0 & 3 & 0 & 3.360127912999990 & -0.928262618399999\\
\hline
0 & 2 & 1 & 0.004300518299999 & -0.002201626699999\\
\hline
0 & 1 & 2 & 0.000000999000000 & -0.000000020900000\\
\hline
0 & 0 & 3 & 0.0 & 0.0\\
\hline
4 & 0 & 0 & 1.045801962500008 & -0.402565733399999\\
\hline
3 & 1 & 0 & -1.420411216799999 & -0.572378273000001\\
\hline
3 & 0 & 1 & 0.003927822499999 & -0.000554230599999\\
\hline
2 & 2 & 0 & -1.122774422700004 & -0.529377973400000\\
\hline
2 & 1 & 1 & 0.000976498800000 & -0.005964260699999\\
\hline
2 & 0 & 2 & 0.0000004915 & 0.000001401500000\\
\hline
1 & 3 & 0 & 0.322474038600001 & 3.267544194000007\\
\hline
1 & 2 & 1 & 0.005696769799999 & -0.001591530000000\\
\hline
1 & 1 & 2 & 0.0000093254 & -0.000008751100000\\
\hline
1 & 0 & 3 & 0.0 & 0.0\\
\hline
0 & 4 & 0 & -3.398481779199986 & 1.658322491499994\\
\hline
0 & 3 & 1 & -0.002802221199999 & 0.001725826300000\\
\hline
0 & 2 & 2 & -0.0000000495 & 0.000000259100000\\
\hline
0 & 1 & 3 & 0.0 & 0.0\\
\hline
0 & 0 & 4 & 0.0 & 0.0\\
\hline
5 & 0 & 0 & -0.831871660400005 & 0.132814269100000\\
\hline
4 & 1 & 0 & 0.949120657999995 & -1.788138425800004\\
\hline
4 & 0 & 1 & -0.003439259299999 & 0.001322541799999\\
\hline
3 & 2 & 0 & -1.903333068200005 & -4.520240236500010\\
\hline
3 & 1 & 1 & -0.005681518600000 & 0.006113348800000\\
\hline
3 & 0 & 2 & 0.0000018543 & 0.000000118100000\\
\hline
2 & 3 & 0 & 5.737533950000011 & 2.994909360000007\\
\hline
2 & 2 & 1 & 0.004100454999999 & -0.003434122199999\\
\hline
2 & 1 & 2 & 0.0000045994 & -0.000004740900000\\
\hline
2 & 0 & 3 & 0.0 & 0.0\\
\hline
1 & 4 & 0 & -4.607335835299982 & 3.560205072699987\\
\hline
1 & 3 & 1 & -0.0017640611 & 0.000632370000000\\
\hline
1 & 2 & 2 & -0.000002936 & 0.000001718100000\\
\hline
1 & 1 & 3 & 0.0 & 0.0\\
\hline
1 & 0 & 4 & 0.0 & 0.0\\
\hline
0 & 5 & 0 & 2.476695333800004 & -1.438056151699995\\
\hline
0 & 4 & 1 & -0.001170340799999 & 0.000608526900000\\
\hline
0 & 3 & 2 & -0.0000001277 & 0.000000238100000\\
\hline
0 & 2 & 3 & 0.0 & 0.0\\
\hline
0 & 1 & 4 & 0.0 & 0.0\\
\hline
0 & 0 & 5 & 0.0 & 0.0\\
\hline
6 & 0 & 0 & 0.250543694400000 & -0.303845693199999\\
\hline
5 & 1 & 0 & -0.872865771300006 & -0.570790623500001\\
\hline
5 & 0 & 1 & -0.0014202491 & -0.000099293799999\\
\hline
4 & 2 & 0 & -2.629992650700005 & -4.544265494000010\\
\hline
4 & 1 & 1 & -0.002726043499999 & 0.004424348300000\\
\hline
4 & 0 & 2 & 0.0000007263 & 0.000000419100000\\
\hline
3 & 3 & 0 & 4.539023046400008 & -1.631924930300001\\
\hline
3 & 2 & 1 & 0.004210737 & 0.000127814799999\\
\hline
3 & 1 & 2 & 0.0000038225 & -0.000004546200000\\
\hline
3 & 0 & 3 & 0.0 & 0.0\\
\hline
2 & 4 & 0 & -2.361741260400014 & 7.959941882000007\\
\hline
2 & 3 & 1 & 0.0004258372 & -0.003437821899999\\
\hline
2 & 2 & 2 & -0.0000105292 & 0.000010411400000\\
\hline
2 & 1 & 3 & 0.0 & 0.0\\
\hline
2 & 0 & 4 & 0.0 & 0.0\\
\hline
1 & 5 & 0 & 2.65929292970001 & -3.122536580600026\\
\hline
1 & 4 & 1 & 0.007337953699999 & -0.008389909299999\\
\hline
1 & 3 & 2 & -0.0000050914 & 0.000004083200000\\
\hline
1 & 2 & 3 & 0.0 & 0.0\\
\hline
1 & 1 & 4 & 0.0 & 0.0\\
\hline
1 & 0 & 5 & 0.0 & 0.0\\
\hline
0 & 6 & 0 & -0.408070357499999 & 0.605073464699998\\
\hline
0 & 5 & 1 & 0.0012935581 & -0.001141694000000\\
\hline
0 & 4 & 2 & -0.0000006819 & 0.000000429700000\\
\hline
0 & 3 & 3 & 0.0 & 0.0\\
\hline
0 & 2 & 4 & 0.0 & 0.0\\
\hline
0 & 1 & 5 & 0.0 & 0.0\\
\hline
0 & 0 & 6 & 0.0 & 0.0\\
\hline
\end{longtable}
\end{center}
\bibliographystyle{plain}
\bibliography{bib}
\end{document}